\documentclass[a4paper]{jpconf}
\usepackage{graphicx}

\newcommand\apj{{\emph{Astrophysical Journal}}}
\newcommand\apjl{{\emph{Astrophysical Journal}}}
\newcommand\apjs{{\emph{Astrophysical Journal, Supplement}}}
\newcommand\aap{{\emph{A\&A}}}
\newcommand\aaps{{A\&AS}}
\newcommand\mnras{{\emph{Monthly Notices of the Royal Astronomical Society}}}
\newcommand\pasj{{\emph{PASJ}}}

\begin{document}
\title{Extragalactic Gamma-ray Background Radiation from Beamed and Unbeamed Active Galactic Nuclei}

\author{Yoshiyuki Inoue}

\address{Department of Astronomy, Kyoto University, Kitashirakawa, Sakyo-ku, Kyoto 606-8502, Japan}

\ead{yinoue@kusastro.kyoto-u.ac.jp}

\begin{abstract}
The origin of the extragalactic gamma-ray background (EGRB) radiation has been a mystery in astrophysics for a long time. Recently the Fermi gamma-ray satellite ({\it Fermi}) has revealed that $\sim$22\% of the unresolved EGRB would be explained by blazars, which are one population of beamed active galactic nuclei (AGNs). The remaining $\sim$78\% of the unresolved EGRB is still unknown. We estimate the contribution of gamma-ray loud radio galaxies, which are misaligned radio loud AGNs recently detected by {\it Fermi}, to EGRB using the radio luminosity function of radio-loud AGNs with the correlation between the radio and gamma-ray luminosities. We find that $\sim$25\% of the unresolved EGRB will be explained by gamma-ray loud radio galaxy population. We also discuss further about the origin of EGRB by comparing the {\it Fermi} EGRB data with our studies on various AGN populations' contribution to EGRB, which are radio quiet AGNs, blazars, and gamma-ray loud radio galaxies.
\end{abstract}

\section{Introduction}
\label{sec:intro}
The origin of the extragalactic gamma-ray background (EGRB) radiation has been argued for a long time in astrophysics. Although several sources have been suggested to explain the MeV and GeV background, it is still discussing about the origin of the MeV background due to the difficulties of MeV gamma-ray measurements. In the case of the GeV background, since blazars which are a type of active galactic nuclei (AGNs) are dominant extragalactic gamma-ray sources \cite{har99,abd10_catalog}, it is expected that unresolved population of blazars would explain the GeV EGRB (see e.g. \cite{ino09}). Recently, \cite{abd10_marco} showed that unresolved blazars can explain $\sim22$\% of the EGRB above 0.1 GeV by analyzing the 11-months {\it Fermi} AGN catalog. Other gamma-ray emitting extragalactic sources have also been discussed as the origin of the GeV EGRB (see \cite{ino11b} for details). 

{\it Fermi} has recently detected GeV gamma-ray emissions from 11 misaligned AGNs (i.e. radio galaxies), which are one type of AGNs with the direction of a relativistic jet {\it not} coinciding with our line of sight \cite{abd10_core}. Their average photon index at 0.1-10 GeV is $\sim2.4$ which is same as that of GeV EGRB \cite{abd10_egrb} and blazars \cite{abd10_marco}. Although they are fainter than blazars, the expected number in the entire sky is much more than blazars. Then, it is naturally expected that they will make a significant contribution to EGRB.

 In this proceeding paper, we discuss the contribution of gamma-ray loud radio galaxies (not blazars) to EGRB based on \cite{ino11b}. Furthermore, we also discuss the EGRB contribution from various AGN populations such as radio-quiet AGNs and blazars based on our previous studies \cite{ino08,ino09}. Throughout this paper, we adopt the standard cosmological parameters of $(h,\Omega_M,\Omega_\Lambda) = (0.7, 0.3, 0.7)$.

\section{Gamma-ray Luminosity Function of Gamma-ray Loud Radio Galaxies}
\label{sec:glf}

To estimate the EGRB contribution of gamma-ray loud radio galaxies, we need to construct a gamma-ray luminosity function (GLF). However, because of a small sample size, it is difficult to construct a GLF using the current gamma-ray data only. Here, the radio luminosity function (RLF) of radio galaxies is extensively studied by previous works (see e.g. \cite{dun90,wil01}). If there is a correlation between the radio and gamma-ray luminosities, we are able to convert the RLF to the GLF with that correlation. 

\begin{figure}[h]
\begin{minipage}{20pc}
\includegraphics[width=90mm]{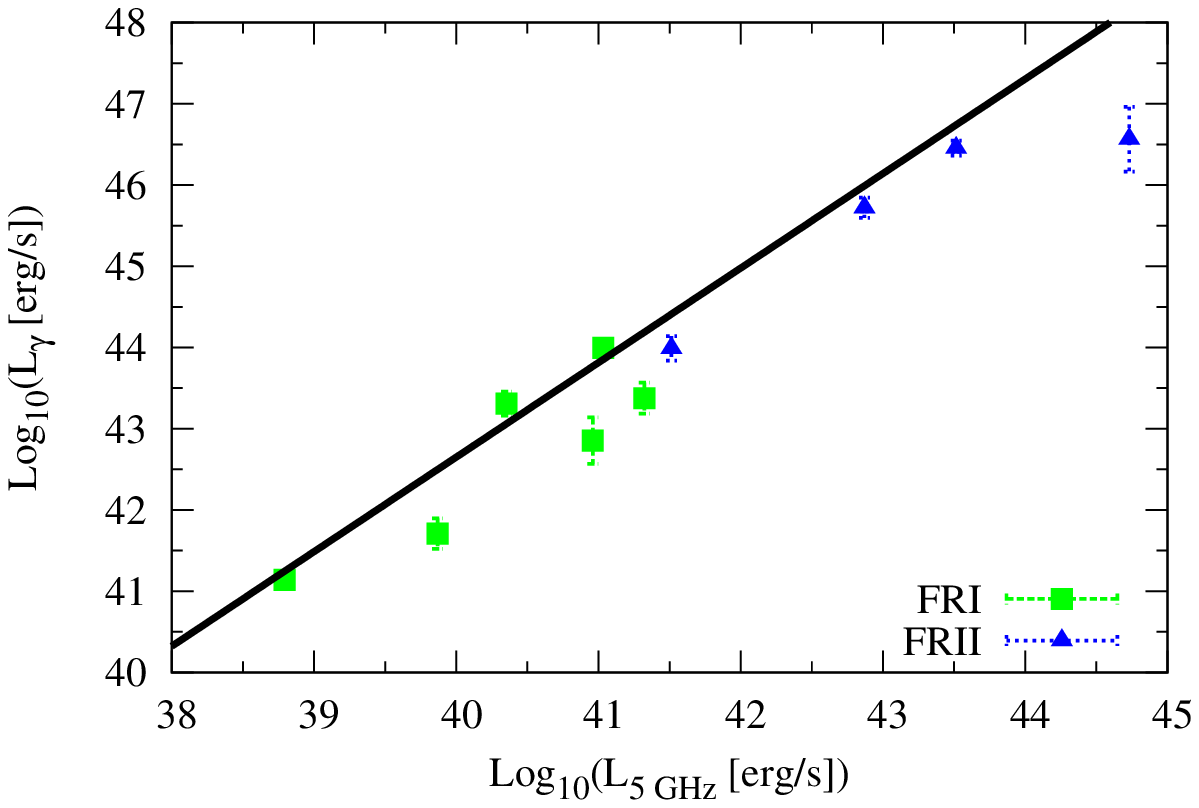}
\caption{\label{fig:lrlg}Gamma-ray luminosity at 0.1--10 GeV versus radio luminosity at 5 GHz. The square and triangle data represents FRI and FRII galaxies, respectively. The solid line is the fit to all sources.}
\end{minipage}\hspace{1pc}%
\begin{minipage}{20pc}
\includegraphics[width=90mm]{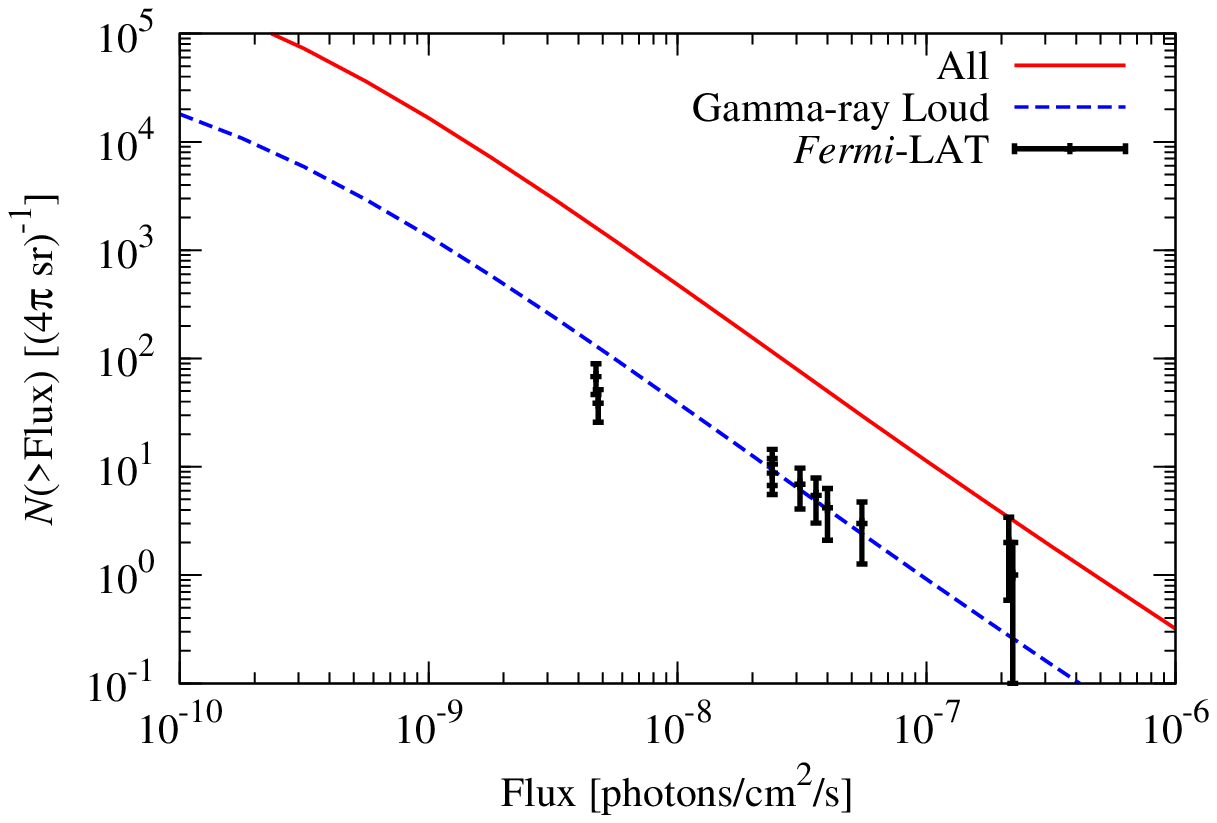}
\caption{\label{fig:logn_logs}The source count distribution of gamma-ray loud radio galaxies in the entire sky. Solid and dashed curve corresponds to the all radio galaxies and gamma-ray loud radio galaxies, respectively. The {\it Fermi} data are also shown.}
\end{minipage} 
\end{figure}

	Figure \ref{fig:lrlg} shows the 5 GHz and 0.1-10 GeV luminosity relation of the {\it Fermi} gamma-ray loud radio galaxies. Square and triangle data represents FRI and FRII radio galaxies, respectively. The solid line shows the fitting line to all the data. The function is given by
\begin{equation}
\log_{10} (L_\gamma) = (-3.90\pm0.61) + (1.16\pm0.02)\log_{10}(L_{\rm 5 GHz}),
\label{eq:lrlg}
\end{equation}
where errors show 1-$\sigma$ uncertainties. In the case of blazars, the slope of the correlation between $L_\gamma(>100 {\rm MeV})$, luminosity above 100 MeV, and radio luminosity at 20 GHz is $1.07\pm0.05$ \cite{ghi10b}. The correlation slopes of gamma-ray loud radio galaxies are similar to that of blazars. This may indicate that emission mechanism is similar in gamma-ray loud radio galaxies and blazars. The partial correlation coefficient \cite{pad92} for these data set becomes 0.866 with chance probability $1.65\times10^{-6}$. Therefore, we conclude that there is a correlation between the radio and gamma-ray luminosities of gamma-ray loud radio galaxies.

	We derive the GLF of gamma-ray loud radio galaxies, $\rho_\gamma(L_\gamma,z)$, using the correlation between the radio and gamma-ray luminosities as Equation \ref{eq:lrlg} and the RLF of radio galaxies, $\rho_r(L_r,z)$, with radio luminosity, $L_r$. The GLF is given as 
\begin{equation}
\rho_\gamma(L_\gamma,z) = \kappa \frac{dL_r}{dL_\gamma}\rho_r(L_r,z),
\label{eq:lf}
\end{equation}
where $\kappa$ is a normalization factor. We use the RLF by \cite{wil01}.

The normalization factor $\kappa$, which corresponds to the fraction of gamma-ray loud radio galaxies against all radio galaxies, is determined by  normalizing our GLF to the source count distribution of the {\it Fermi} radio galaxies, which is sometimes called logN--logS plot or cumulative flux distribution. Source count distribution is calculated by
\begin{equation}
N(>F_\gamma)=4\pi\int_{0}^{z_{\rm max}}dz \frac{d^2V}{d\Omega dz}\int_{L_\gamma(z,F_\gamma)}^{L_{\gamma,\rm max}}dL_\gamma\rho_\gamma(L_\gamma,z),
\end{equation}
where $L_\gamma(z,F_\gamma)$ is the gamma-ray luminosity of a blazar at redshift $z$ whose photon flux at $>$100 MeV is $F_\gamma$. Hereinafter, we assume $z_{\rm max}=5$ and $L_{\gamma,\rm max}=10^{48} {\rm erg/s}$ in this study. These assumptions hardly affect the results in this study.

Since the completeness of the {\it Fermi} sky survey depends on the photon flux and photon index of a source, we need to take into account this effect (so called the detection efficiency) to compare GLF with the cumulative source count distribution of gamma-ray loud radio galaxies. The detection efficiency of {\it Fermi} is shown in Figure 7 of \cite{abd10_marco}.

Figure \ref{fig:logn_logs} shows the source count distribution of gamma-ray loud radio galaxies. The data is after the conversion of the detection efficiency. Solid red curve shows in the case of $\kappa=1$ which corresponds to the case that all radio galaxies emit gamma-rays. Dashed blue curve corresponds to the GLF fitted to the {\it Fermi} data with $\kappa = 0.081\pm0.011$. $\sim1000$ gamma-ray loud radio galaxies are expected with 100\% complete entire sky survey above the flux threshold $F_{\gamma}(>100 {\rm MeV})=1.0\times10^{-9}\ {\rm photons\ cm^{-2}\ s^{-1}}$ above 100 MeV. We note that the current detection efficiency of {\it Fermi} at $F_{\gamma}(>100 {\rm MeV})=1.0\times10^{-9}\ {\rm photons\ cm^{-2}\ s^{-1}}$ is $\sim10^{-3}$.

\section{EGRB from Gamma-ray Loud Radio Galaxies}
\label{sec:egrb}

We calculate the EGRB spectrum by integrating our GLF in the redshift and luminosity space, using the a simple broken power law SED model (see \cite{ino11b} for details). The EGRB spectrum is calculated as 
 
 \begin{equation}
\frac{d^2F(\epsilon)}{d\epsilon d\Omega} =\frac{c}{4\pi}\int_0^{z_{\rm max}}dz\left|\frac{dt}{dz}\right|\int_{L_{\gamma,\rm min}}^{L_{\gamma,\rm max}}dL_\gamma \rho_\gamma(L_\gamma,z) \frac{dL[L_\gamma,(1+z)\epsilon]}{d\epsilon}\exp[-\tau_{\gamma,\gamma}(\epsilon,z)]\times \{1.0 - \omega(F_\gamma[L_\gamma,z])\} ,
\end{equation}
where $t$ is the cosmic time and $dt/dz$ can be calculated by the Friedmann equation in the standard cosmology. The minimum gamma-ray luminosity is set at $L_{\gamma,\rm min} = 10^{39} {\ \rm erg/s}$. $\omega(F_\gamma[L_\gamma,z])$ is the detection efficiency of {\it Fermi} at the photon flux $F_\gamma(L_\gamma,z)$ .

Here, high energy $\gamma$-rays propagating the universe are absorbed by the interaction with the extragalactic background light (EBL), also called as cosmic optical and infrared background \cite{kne04,fin10}. $\tau_{\gamma,\gamma}(\epsilon,z)$ is the optical depth of this background radiation. In this study, we adopt the EBL model by \cite{fin10}. We also include the contribution of the cascade emission. The gamma-ray absorption creates electron--positron pairs. These pairs scatter the cosmic microwave background radiation to make the secondary emission component (so called cascade emission) to the absorbed primary emission \cite{aha94}.
  
\begin{figure}[h]
\includegraphics[width=100mm]{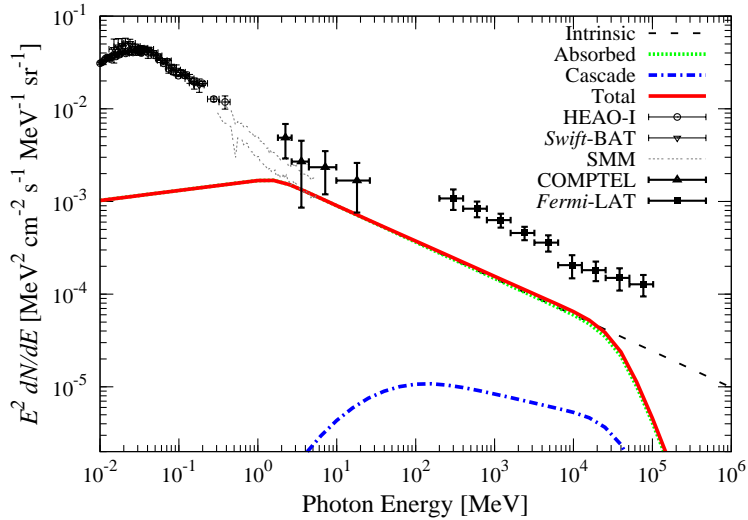}\hspace{1pc}%
\begin{minipage}[b]{14pc}\caption{\label{fig:egrb}The unresolved EGRB spectrum from gamma-ray loud radio galaxies. Dashed, dotted, dot-dashed, and solid curves show the intrinsic spectrum (no absorption), absorbed, cascade, and total (absorbed+cascade) EGRB spectrum, respectively. The observed data of {\it HEAO-I} \cite{gru99}, {\it Swift}-BAT \cite{aje08}, {\it SMM} \cite{wat97}, COMPTEL \cite{kap96}, and {\it Fermi}-LAT \cite{abd10_egrb} are also shown by the symbols indicated in the figure.}
\end{minipage}
\end{figure}

 Fig. \ref{fig:egrb} shows the unresolved $\nu I_\nu$ EGRB spectrum of gamma-ray loud radio galaxies in the unit of ${\rm MeV^2cm^{-2}s^{-1} MeV^{-1}sr^{-1}}$ predicted by our GLF. The intrinsic (the spectrum without the EBL absorption effect), absorbed, and cascade components of the EGRB spectrum and the total EGRB spectrum (absorbed+cascade) are shown. As in the figure, the cascade emission does not contribute to the EGRB spectrum significantly.

The expected EGRB photon flux above 100 MeV from gamma-ray loud radio galaxy populations is $0.26\times10^{-5}\ {\rm photons \ cm^{-2}\ s^{-1}\ sr^{-1}}$. As the unresolved {\it Fermi} EGRB flux above 100 MeV is $1.03\times10^{-5}\ {\rm photons \ cm^{-2}\ s^{-1}\ sr^{-1}}$ \cite{abd10_egrb}, the gamma-ray loud radio galaxies explains $\sim25$\% of the unresolved EGRB flux. For the comparison, recent analysis of {\it Fermi} blazars showed that blazars explains $\sim22$\% of the unresolved EGRB \cite{abd10_marco}. Therefore, we are able to explain $\sim47$\% of EGRB by radio-loud AGN populations. We note that, because of small samples and the uncertainty in the correlation between radio and gamma-ray luminosities, the fraction of gamma-ray loud radio galaxies in unresolved EGRB varies from $\sim10$\% to $\sim63$\%. 

\section{EGRB from Beamed and Unbeamed AGNs}
\begin{figure}[h]
\begin{center}
\includegraphics[width=120mm]{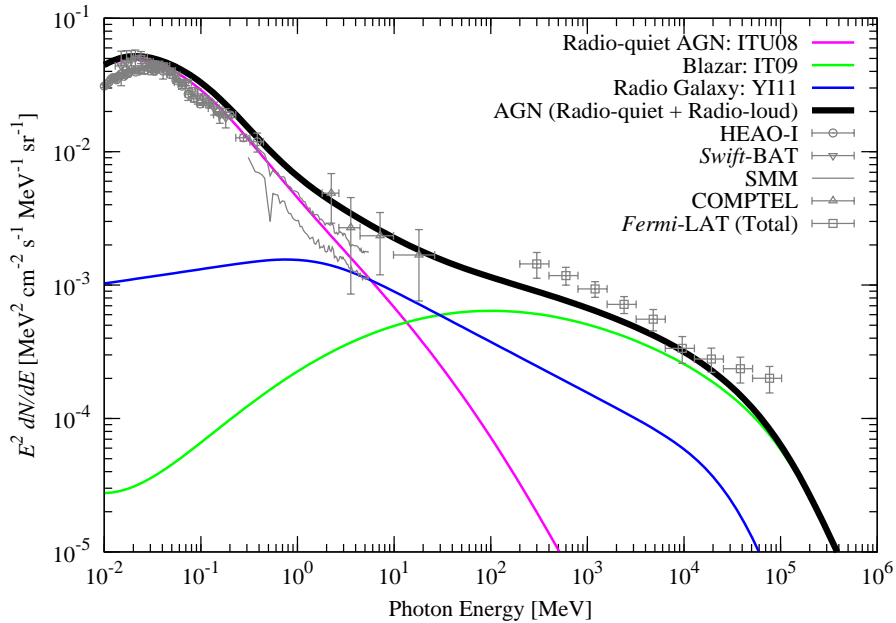}
\caption{\label{fig:egrb_total}EGRB spectrum from various AGN populations. Magenta, green, blue, and black curves show the EGRB contribution from radio-quiet AGN \cite{ino08}, blazars \cite{ino09}, gamma-ray loud radio galaxies \cite{ino11b}, and total AGN populations (radio-quiet AGNs, blazars, and gamma-ray loud radio galaxies), respectively. The observed data are same as Fig. \ref{fig:egrb}, but total (resolved + unresolved) EGRB data are shown here.}
\end{center}
\end{figure}

We have also conducted works on EGRB from radio-quiet AGNs \cite{ino08} and blazars \cite{ino09}. Raio-quiet AGNs would significantly contribute to the MeV background by including non-thermal electron population in the AGN hot corona above the accretion disk \cite{ino08}. Our blazar GLF model \cite{ino09} is constructed based on the EGRET data taking into account blazar sequence \cite{fos98,kub98}. This blazar GLF is consistent with various {\it Fermi} data (source counts and EGRB contribution) \cite{ino10a,ino11}. Fig. \ref{fig:egrb_total} shows the EGRB contribution of radio-quiet AGNs, blazars, and gamma-ray loud radio galaxies (i.e. beamed and unbeamed AGN populations). As in the figure, AGN populations are dominant in the cosmic background radiation from X-ray to gamma-ray. In this figure, we use the total (resolved + unresolved) EGRB data to avoid the instrument dependence, since the model by \cite{ino09} is based on the EGRET data.
   
\section{Origin of the GeV EGRB}
\label{sec:dis}

\begin{figure}[h]
\includegraphics[width=100mm]{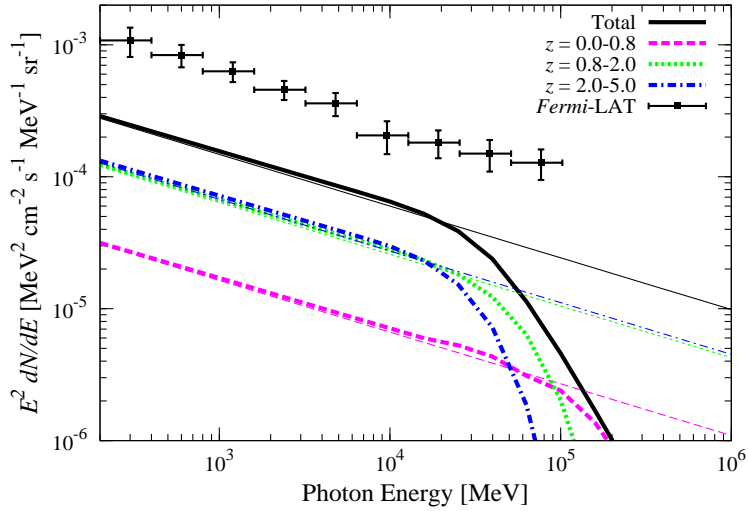}\hspace{1pc}%
\begin{minipage}[b]{16pc}\caption{\label{fig:egrb_z}Gamma-ray loud radio galaxy EGRB spectra  at each redshift bins. Solid, dashed, dotted, and dot-dashed curves shows the EGRB spectrum at $z=0.0-5.0$, $z=0.0-0.8$, $z=0.8-2.0$, and $z=2.0-5.0$, respectively. The thin and thick curve corresponds to the intrinsic spectrum (not absorbed) and the total spectrum (absorbed + cascade), respectively. The unresolved EGRB data of {\it Fermi}-LAT \cite{abd10_egrb} is also shown.}
\end{minipage}
\end{figure}

Although AGN populations dominate EGRB, it is difficult to explain 100\% of EGRB. Fig. \ref{fig:egrb_z} shows the unresolved gamma-ray loud radio galaxy EGRB spectra at each redshift bins. Because of EBL, the spectrum above $>30$ GeV shows the absorbed signature. Here, the cosmological sources have their evolution peaks at $z=1\sim2$ such as cosmic star formation history and AGN activity (see e.g. \cite{hop06,ued03}). This means that the gamma-rays from extragalactic sources (e.g. galaxies and AGNs) will experience the EBL absorption. However, as shown in Fig. \ref{fig:egrb_z}, the {\it Fermi} EGRB spectrum does not clearly show such an absorbed signature. This might suggest that nearby gamma-ray emitting sources or sources with very hard gamma-ray spectrum would be the dominant population of EGRB above 10 GeV such as high-frequency-peaked BL Lacs. To address this issue, we should await the EGRB information above 100 GeV by future observations such as {\it Fermi}. CTA would also be possible to see the EGRB at much higher energy band. We also need to examine the EBL models at high redshift. It is expected that CTA will see blazars up to $z\sim1.2$ at very high energy band $>30$ GeV \cite{ino10a}. Therefore, {\it Fermi} and CTA will be a key to understanding the origin of EGRB. 

\section{Conclusion}
\label{sec:con}
In this paper, we studied the contribution of gamma-ray loud radio galaxies to the EGRB by constructing their GLF. First, we explored the correlation between the radio and gamma-ray luminosities of gamma-ray loud radio galaxies which are recently reported by {\it Fermi}. We found that there is a correlation $L_\gamma\propto L_{\rm 5GHz}^{1.16}$ by a partial correlation analysis.

Based on this correlation, we defined the GLF of gamma-ray loud radio galaxies using the RLF of radio galaxies. We normalized the GLF to fit to the cumulative flux distribution of {\it Fermi} samples. Then, we predicted the contribution of gamma-ray loud radio galaxies to the MeV and GeV EGRB. We found that gamma-ray loud radio galaxies will explain $\sim25$\% of the EGRB flux above 100 MeV and also make a significant contribution to the 1--30 MeV EGRB. Since blazars explain $\sim22$\% of EGRB, $\sim47$\% of EGRB would be explained by blazars and gamma-ray loud radio galaxies.

Based on our previous studies of radio-quiet AGNs and blazars, we also showed the contribution of beamed and unbeamed AGN populations to EGRB. AGN populations dominate the cosmic X-ray and gamma-ray background in about 7 orders magnitude of photon energy.

We also make an interpretation on the origin of the EGRB above 10 GeV from the point of view of the EBL absorption effect. If the EBL absorption signature is still not appeared in the EGRB spectrum, the origin would be nearby sources or sources with hard gamma-ray spectrum. We should await the EGRB data at higher energy band for this issue.

\ack The author thanks the organizing committees of the conference {\it Beamed and Unbeamed Gamma-rays from Galaxies} at Lapland Hotel Olos, Muonio, Finland for providing great opportunities of lots of discussions. The author acknowledges support by the Research Fellowship of the Japan Society for the Promotion of Science (JSPS). This work was supported in part by
the Global COE Program ``The Next Generation of Physics, Spun from
Universality and Emergence'' at Kyoto University from the Ministry of Education, Culture,
Sports, Science and Technology (MEXT) of Japan.

\end{document}